\documentclass[a4paper,10pt]{article}

\usepackage{contribution1}

\newcommand{\weblink}[2][]{%
    \ifthenelse{\equal{#1}{}}%
    {\textnormal{\url{#2}}}%
    {\textnormal{\href{#2}{#1}}}%
}

\def\beq{\begin{equation}}
\def\eeq#1{\label{#1}\end{equation}}
\def\eeqn{\end{equation}}

\def\beqa{\begin{eqnarray}}
\def\eeqa#1{\label{#1}\end{eqnarray}}
\def\eeqan{\end{eqnarray}}

\let\bar=\overbar

\def\D{{\cal D}}

\def\Dslash{\not{\hbox{\kern-4pt $D$}}}
\def\dslash{\not{\hbox{\kern-2pt $\del$}}}

\def\msb{{\bar{\ssstyle M \kern -1pt S}}}

\newcommand{\contribution}[7][]{%
  \clearpage
  \thispagestyle{plain}

  \ifthenelse{\equal{#1}{}}
  {\hypersetup{pdftitle={#2}}}
  {\hypersetup{pdftitle={#1}}}
  \hypersetup{pdfauthor={{#3} {#4}}}
  {\centering\normalfont\LARGE\bfseries\sffamily #2 \par\nobreak}
  \lhead{}
  \chead{%
    \textit{\footnotesize XXIInd International Workshop ``High-Energy Physics and Quantum Field Theory'',
June 24 -- July 1, 2015, Samara, Russia}%
  }
  \rhead{}
  \bigskip
  \begin{center}
    {#3} {#4}\ifthenelse{\equal{#6}{}}{}{\footnote{\weblink[#6]{mailto:#6}}}
    \ifthenelse{\equal{#7}{}}{}{#7} \\
    \textit{#5}
  \end{center}
  \bigskip
}

\renewcommand{\abstract}[1]{%
  \begin{center}
    \begin{minipage}{0.85\textwidth}
      \begin{footnotesize}
        #1
      \end{footnotesize}
    \end{minipage}
  \end{center}
  \bigskip
}

\begin{document}

%
%
%
%
%
%
{  


%

\contribution[Static electromagnetic moments and lepton decay
constant of the $\rho$-meson]  
{Static electromagnetic moments and lepton decay constant of the $\rho$-meson in the instant form of relativistic quantum mechanics}  
{Alexander}{Krutov,} 
{Samara State University, \\
Academician Pavlov St., 1, 443011 Samara, Russia\\D.V.~Skobeltsyn
Institute of Nuclear Physics,
M. V. Lomonosov Moscow State University, 119991 Moscow, Russia} 
{krutov@samsu.ru}
{Roman Polezhaev, Vadim Troitsky}  

%

\abstract{%
The relativistic calculations of the electromagnetic form factors,
static moments and lepton decay constant of the $\rho$-meson are
given in the framework of the instant form of the relativistic
quantum mechanics  with different model wave functions. The
modified impulse approximation (MIA) is used for the electroweak
current operator. MIA provides Lorentz covariance and conservation
law for current operators. The developed approach gives the nonzero
quadrupole moment due to the relativistic Wigner spin rotation in
the $S$-state of the quarks in $\rho$-meson.}
%

\section{Introduction}

The study of the electromagnetic structure of the $\rho$- meson
remains relevant for many years. Interest in the study of
electroweak structure of the $\rho$- meson has risen again
recently in connection with new experimental data for the
corresponding leptonic decay constant: $f^{exp}_{\rho}=152\pm 8
MeV$\cite{Oli14}. This result is very important
$\rho$-meson is short-lived particle and the experimental
information about its electroweak properties (electromagnetic form
factors and related static moments, charge radii, etc.) is poorly
enough.  However, there are a large number of theoretical approaches
and models for description of this meson
\cite{ChC04,BhM08,AlO09,RoB11,Sim13,BiS14,MeS15}.

In this paper the description of the electromagnetic static
moments and lepton decay constant of the $\rho$-meson is
presented. Our approach is based on the use of the instant form
(IF) of relativistic quantum mechanics (RQM). The detailed
description of RQM can be found in the review \cite{KeP91}.

The basic point of our approach  is the procedure of the
construction of the electroweak current operator
\cite{KrT02,KrT03prc}. This construction is from the
general method of relativistic invariant parameterization of local
operator matrix elements proposed in the Ref. \cite{ChS63}.

In fact, this procedure is a realization of the Wigner--Eckart
theorem for the Poincar\'e group and it enables one for given
matrix element of arbitrary tensor dimension to separate the
reduced matrix elements (form factors) that are invariant under
the Poincar\'e group. The matrix element of a given operator is
represented as a sum of terms, each one of them being a covariant
part multiplied by an invariant part. In such a representation a
covariant part describes transformation (geometrical) properties
of the matrix element, while all the dynamical information on the
transition is contained in the invariant part -- reduced matrix
elements or form factors.

In our approach some rather general problems arising in the
constituent quark models have been solved. For example, our
description of composite systems, in fact, solves the problem of
construction of the electromagnetic current satisfying the
conditions of Lorentz covariance and conservation law
\cite{KrT02}.

Our calculations of the electroweak characteristics of
$\rho$-meson are performed in the well known impulse approximation
(IA). It means that the electromagnetic current of a composite
system is a sum of one--particle currents of the constituents. It
is worth emphasizing that in our method this approximation does
not violate the standard conditions for the current. This is a
variant of the relativistic impulse approximation formulated in
terms of reduced matrix elements  -- modified impulse
approximation (MIA) (see, e.g., \cite{KrT02}).

Using different model wave functions of quark in meson we
calculate the electromagnetic form factors, the static properties
and the lepton decay constant of $\rho$-meson supposing quarks to
be in the $S$-state of relative motion. It is interesting to
mention that relativistic effects occur to produce a nonzero
quadrupole moment and quadrupole form factor. It is well known
that in the nonrelativistic case the non--zero quadrupole form
factor is caused by the presence of the $D$- wave and is zero
otherwise.

\section{Integral representation for the
$\rho$-meson electromagnetic form factors and leptonic decay constant}

Let us consider the matrix element of the $\rho$-meson
electromagnetic current. In our constituent quark model the $u$-
and $\bar d$-- quarks are in the $S$-state of relative motion,
that is $l = l' =$ 0, with following total spin and total angular
momentum: $J=J'=S=S'=$1. This matrix element is given in the Ref.
\cite{KrT03prc}:
\begin{equation}
\langle\vec p_c\,,m_{J\rho}|j_\mu(0)|\vec p_\rho\,'\,,m'_{J\rho}\rangle
= \langle\,m_{J\rho}|\,D^{1}(p_\rho\,,p'_\rho)\,
\sum_{i=1,3}\,
\tilde{\cal F}\,^i_\rho(t)\,\tilde A^i_\mu\,|m'_{J\rho}\rangle\;;
\label{jc=FA}
\end{equation}
$$
\tilde{\cal F}\,^1_c(t) = \tilde f^\rho_{10} +
\tilde f^\rho_{12}\left\{[i{p_\rho}_\nu\,\Gamma^\nu(p'_\rho)]^2
\right.
\left. - \frac{1}{3}\,\hbox{Sp}[i{p_\rho}_\nu\,\Gamma^\nu(p'_\rho)]^2\right\}
\frac{2}{\hbox{Sp}[{p_\rho}_\nu\,\Gamma^\nu(p'_\rho)]^2}\;,
$$
\begin{equation}
\tilde{\cal F}\,^3_\rho(t) = \tilde f^\rho_{30}\;;
\label{FicAic}
\end{equation}
$$
\tilde A^1_\mu = (p_\rho + p'_\rho)_\mu\;,\quad
\tilde A^3_\mu =
\frac{i}{M_\rho}
\varepsilon_{\mu\nu\lambda\sigma}
\,p_\rho^\nu\,p'_\rho\,^\lambda\,\Gamma^\sigma(p'_\rho)\;.
$$
here $p'_\rho, p_\rho$ are 4-momentum of  $\rho$-meson in initial
and final states, respectively, $m'_{J\rho}\,,m_{J\rho}$ are
projections of the total angular momenta,
$D^{1}(p_\rho\,,\,p'_\rho)$ is matrix of Wigner rotation,
$\Gamma^\nu(p'_\rho)$ is 4-vector of spin, $M_\rho$ is
$\rho$-meson mass, $\tilde f^\rho_{10}\,,\,\tilde f^\rho_{12}\,,$
$\tilde f^\rho_{30}$ are charge, quadrupole and magnetic form
factors of $\rho$-meson, respectively.

The integral representations for the composite system form factors
in (\ref{jc=FA}), and (\ref{FicAic}) are obtained in the Refs.
\cite{KrT02,KrT03prc}:
\begin{equation}
\tilde f^\rho_{in}(Q^2) =  \int\,d\sqrt{s}\,d\sqrt{s'}\,
\varphi(s)\,\tilde G_{in}(s,Q^2,s')\varphi(s')\;,
\label{intrepJ1}
\end{equation}
here $\varphi(s)$ is wave function of quarks in RQM, $\tilde
G_{in}(s,Q^2,s')$ is Lorentz-covariant generalized function
(reduced matrix element on the Poincar\'e group).

Our form factors in (\ref{FicAic}) can be written in terms of
conventional Sachs form factors for the system with the total
angular momentum equal 1.  To do this let us write the
parameterization of the electromagnetic current matrix element in
the Breit frame (see, e.g.~\cite{ArC80}):
$$
\langle\vec p_\rho\,,m_J|j_\mu(0)|\vec p_\rho\,'\,,m'_J\rangle =
G^\mu(Q^2)\;,
$$
$$
G^0(Q^2) = 2p_{\rho0}\left\{(\vec\xi\,'\vec\xi\,^*)\,G_C(Q^2) +
\left[(\vec\xi\,^*\vec Q)(\vec\xi\,'\vec Q) - \frac{1}{3}Q^2
(\vec\xi\,'\vec\xi\,^*)\right]\,\frac{G_Q(Q^2)}{2M_\rho^2}\right\}\;,
$$
\begin{equation}
\vec G(Q^2) = \frac{p_{\rho0}}{M_\rho}\left[\vec\xi'\,(\vec\xi\,^*\vec Q) -
\vec\xi\,^*(\vec\xi\,'\vec Q)\right]\,G_M(Q^2)\;.
\label{Gi}
\end{equation}
Here $G_C\;,\;G_Q\;,\;G_M$  are the charge, quadrupole and
magnetic form factors, respectively.

The polarization vector in the Breit frame has the
following form:
\begin{equation}
\xi^\mu(\pm 1) = \frac{1}{\sqrt{2}}(0\;,\;\mp 1\;,\;-\,i\;,\;0)\;,\quad
\xi^\mu(0) = (0\;,\;0\;,\;0\;,\;1)\;.
\end{equation}
The variables in $\xi$ are total angular momentum
projections.

In the Breit frame:
\begin{equation}
q^\mu = (p_\rho - p'_\rho)^\mu = (0\;,\;\vec Q)\;,\quad
p_\rho^\mu = (p_{\rho0}\;,\;\frac{1}{2}\vec Q)\;,\quad
p'_\rho\,^\mu = (p_{\rho0}\;,\;-\frac{1}{2}\vec Q)\;,
\end{equation}
$$
p_{\rho0} = \sqrt{M_\rho^2 + \frac{1}{4}Q^2}\;,\quad
\vec Q = (0\;,\;0\;,\;Q)\;.
$$
Comparing (\ref{jc=FA}) and (\ref{Gi}) and taking into account the
fact that in the Breit system $D^1_{m_J\,m''_J}(p_\rho\,,p_\rho')
= \delta_{m_J\,m''_J}\;,$ we have:
$$
G_C(Q^2) = \tilde f^\rho_{10}(Q^2)\;,\quad
G_Q(Q^2) = \frac{2\,M_\rho^2}{Q^2}\,\tilde f^\rho_{12}(Q^2)\;,
$$
\begin{equation}
\quad G_M(Q^2) = -\,M_\rho\,\tilde f^\rho_{30}(Q^2)\;.
\label{GF}
\end{equation}

Let us use for (\ref{intrepJ1}) the modified impulse approximation
formulated in terms of form factors $\tilde G_{iq}(s,Q^2,s')$. The
physical meaning of this approximation is considered in detail in
Ref. \cite{KrT02}. In the frame of MIA the invariant form factors
$\tilde G_{iq}(s,Q^2,s')$ in (\ref{intrepJ1}) are changed by the
so called  free two--particle invariant form factors
$g_{0i}(s,Q^2,s')\;(i=C,Q,M)$ describing the electromagnetic
properties of the system of two free particles with the quantum
numbers of the $\rho$-meson. So, the equations to be used for the
calculation of the $\rho$--meson electromagnetic properties in MIA
are the following:
$$
G_C(Q^2) =
\int\,d\sqrt{s}\,d\sqrt{s'}\, \varphi(s)\,g_{0C}(s\,,Q^2\,,s')\,
\varphi(s')\;,
$$
\begin{equation}
G_Q(Q^2) =
\frac{2\,M_\rho^2}{Q^2}\,\int\,d\sqrt{s}\,d\sqrt{s'}\,
\varphi(s)\,g_{0Q}(s\,,Q^2\,,s')\,\varphi(s')\;,
\label{GqGRIP}
\end{equation}
$$
G_M(Q^2) =-\,M_\rho\,\int\,d\sqrt{s}\,d\sqrt{s'}\,
\varphi(s)\,g_{0M}(s\,,Q^2\,,s')\,
\varphi(s')\;.
$$

The free two--particle invariant form factors can be calculated by
the methods of relativistic kinematics and have the form given in
Ref. \cite{KT03}.

For the calculation of  the $\rho$-meson leptonic decay constants
we used the method of parametrization matrix element of the
current which is nondiagonal with respect to the total angular
momentum \cite{KrP15}.

Leptonic decay constant of the vector meson $f_{\rho}$ is
determined by the following matrix element of the electroweak
current (see, e.g. \cite{Jau03}):
\begin{equation}\label{const11}
\langle0|j^\rho_{\mu}(0)|\vec{P_{\rho}},m_{\rho}\rangle =
i\,\sqrt{2}f_{\rho}\,\xi_{\mu}(m_\rho)\frac{1}{(2\pi)^{3/2}}\;,
\end{equation}

In MIA the lepton decay constant of the $\rho$-meson is given by
expression:
\begin{equation}
f_\rho =
\int\,d\sqrt{s}\,G^{1,0,1}_{0,1}(s)\,\varphi(s)\;,\label{fc4-fi}
\end{equation}
where the corresponding free form factor $G^{1,0,1}_{0,1}(s)$
is:
\begin{equation}
G^{1,0,1}_{0,1}(s) =-\frac{3\sqrt{3}(\sqrt{s}+2M)}{16\sqrt{s}\pi^2}
\left(\frac{7s+12M\sqrt{s}+8M^2}{6s+12M\sqrt{s}+12M^2}\right)\;.
\label{freeff}
\end{equation}

Thus an analytic expression for the constants can be represented as follows:
\begin{eqnarray}\label{decayconst111}
&&\hspace{-2mm}f_{\rho}=\frac{\sqrt{3}}
{\pi \sqrt{2}}\int^{\infty}_{0}\,dk\,k^2\,\psi(k)
\frac{(\sqrt{k^2 + M^2} +
M)}{(k^2 + M^2)^{3/4}}\nonumber\\
[2mm]&& \times\, \left(1 +
\frac{k^2}{3(\sqrt{k^2+M^2}+M)^2}\right)\;.
\end{eqnarray}
Note that an analytical expression for the $\rho$-meson leptonic
decay constants obtained in our approach coincides with Refs.
\cite{And99,Jau03}.

\section{The electromagnetic structure of the $\rho$-meson}

In this section we make use of the results of the previous
sections to calculate the $\rho$-meson electromagnetic properties.

The $\rho$-meson electromagnetic form factors are calculated using
(\ref{GqGRIP}) in MIA. The wave functions in sense of RQM in
(\ref{GqGRIP}) at $J=S=1\;,l=0$ are defined by the following
expression (see, e.g. \cite{KrT02}):
\begin{equation}
\varphi(s) =\sqrt[4]{s}\,\psi(k)\,k\;,\quad k = \frac{1}{2}\sqrt{s
- 4\,M^2}\;,
\label{phi(s)}
\end{equation}
and is normalized by the condition:
\begin{equation}
\int\,\psi^2(k)\,k^2\,dk = 1\;,
\label{norm}
\end{equation}
here $\psi(k)$ is a model wave function.

For the description of the relative motion of quarks the following phenomenological wave functions
are used:

1. A Gaussian or harmonic oscillator wave function (HO) (see, e.g.
\cite{CoP05}):
\begin{equation}
\psi(k)= N_{HO}\,
\hbox{exp}\left(-{k^2}/{2\,b^2}\right).
\label{HO-wf}
\end{equation}

2. A power-law wave function (PL) (see, e.g. \cite{CoP05}):
\begin{equation}
\psi(k) =N_{PL}\,{(k^2/b^2 +
1)^{-n}}\;,\quad n = 2\;,3\;.
\label{PL-wf}
\end{equation}

For Sachs form factors of quarks we have:
\begin{equation}
G^{q}_{E}(Q^2) =
e_q\,f_q(Q^2)\;,\quad G^{q}_{M}(Q^2) = (e_q + \kappa_q)\,f_q(Q^2)\;,
\label{q ff}
\end{equation}
where $e_q$ is the quark charge and $\kappa_q$ is the quark
anomalous magnetic moment.  For $f_q(Q^2)$ the form proposed in
Ref. \cite{KrT98} is used:
\begin{equation}
f_q(Q^2) =
\frac{1}{1 + \ln(1+ \langle r^2_q\rangle Q^2/6)}\;.
\label{f_qour}
\end{equation}
Here $\langle r^2_q\rangle$ is the mean square radius (MSR) of
constituent quark.

The reasons of the choice for the function $f_q(Q^2)$ can be found
in Ref. \cite{KrT98} (see also Ref.\cite{KrT02}) The form
(\ref{f_qour}) is based on the fact that this expression gives
the asymptotics of the pion form factor at $Q^2\to\infty$ which
coincides with the QCD prediction (see, e.g.\cite{MaM73}).

So, for the calculations we use a conventional set of parameters
of constituent quark model. The structure of the constituent quark
is described by the following parameters: $M_u = M_{\bar d} = M$
is the constituent quark mass, $\kappa_u\;,\;\kappa_{\bar d}$ are
the constituent quarks anomalous magnetic moments,
$\langle\,r^2_u\rangle = \langle\,r^2_{\bar d}\rangle =
\langle\,r^2_q\rangle$ is the quark MSR. The interaction of quarks
in $\rho$ meson is characterized by wave functions (\ref{HO-wf})
-- (\ref{PL-wf}) with the parameters $b$.

Since the quark composition of the pion and $\rho$-meson are the
same, then the parameters were fixed in our calculation as follows
(see Ref.\cite{KrT01}).

1. We use $M$=0.22 GeV  for the quark mass; the quark anomalous
magnetic moments enter the our expressios through the sum
($\kappa_u + \kappa_{\bar d}$) and we take $\kappa_u +
\kappa_{\bar d}$ = 0.0268 in natural units; for the quark MSR we
use the relation $\langle r^2_q\rangle \simeq 0.3 /M^2$.

2. The parameter of the wave function $b$ was fixed from
the requirements of the description of experimental values of
$\rho$-meson leptonic decay constants and following relation for
the $\rho$-meson MSR from refs. \cite{CaG96,VoL90}:
\begin{equation}
\label{Wu}
\langle r^2_\rho\rangle - \langle r^2_\pi\rangle = 0.11\pm 0.06
fm^2\;.
\end{equation}

For the pion MSR the experimental data is taken from Ref.
\cite{Oli14}: $\langle r^2_\pi\rangle^{1/2}$ = 0.672$\pm$0.008 fm.

The $\rho$-meson MSR is calculated from relation:
\begin{equation}
\langle\,r^2_\rho\rangle = -6\,G'_{C}(0)\;.
\label{MSR}
\end{equation}

The $\rho$-meson mass in (\ref{GqGRIP}) is taken from Ref.
\cite{Oli14}:  $M_\rho$ = 775.5$\pm$0.4 MeV.

The magnetic $\mu_\rho$ and the quadrupole $Q_\rho$ moments of
$\rho$ meson were calculated using the relations given in Ref.
\cite{ArC80}:
\begin{equation}
G_M(0) = \frac{M_\rho}{M}\,\mu_\rho\;,\quad
G_Q(0) = M_\rho^2\,Q_\rho\;.
\label{stat}
\end{equation}
The static limit in (\ref{GqGRIP}) gives the following
relativistic expressions for moments:
\begin{equation}
\mu_\rho = \frac{1}{2}\,\int_{2M}^\infty\,d\sqrt{s}\,\frac{\varphi^2(s)}
{\sqrt{s - 4\,M^2}}\,\left\{1 - L(s) + (\kappa_u + \kappa_{\bar d})
\left[1 - \frac{1}{2}\,L(s)\right]\right\}\;,
\label{mu}
\end{equation}
\begin{equation}
Q_\rho = -\,\int_{2M}^\infty\,d\sqrt{s}\,\frac{\varphi^2(s)}
{\sqrt{s}}\,
\left[\frac{M}{\sqrt{s} + 2\,M} + \kappa_u + \kappa_{\bar d}\right]
\,\frac{L(s)}{4M\sqrt{s - 4M^2}}\;,
\label{Q}
\end{equation}
$$
L(s) = \frac{2\,M^2}{\sqrt{s - 4\,M^2}\,(\sqrt{s} + 2\,M)}\,\left[
\frac{1}{2\,M^2}\sqrt{s\,(s - 4\,M^2)} + \right.
$$
$$
+ \left.
\ln\,
\frac{\sqrt{s} - \sqrt{s - 4\,M^2}}{\sqrt{s} + \sqrt{s - 4\,M^2}}\right]\;.
$$
Let us note that the nonzero $\rho$-meson quadrupole moment
appears due to the relativistic effect of Wigner spin rotation of
quarks only. So, measuring of this quadrupole moment can be a test
of the relativistic invariance in the confinement region.

\begin{table} [h!]
\caption{The $\rho$-meson electromagnetic moments and lepton decay
constant obtained with the different model wave functions
\protect(\ref{HO-wf}) -- \protect(\ref{PL-wf}).
$\langle\,r^2_{\rho}\rangle$ is MSR in $fm^2$, $\mu_\rho$ is
relativistic magnetic moment (\ref{mu}) in natural units, $Q_\rho$
is quadrupole moment (\ref{Q}) in fm$^2$. The parameters $b$ in
\protect(\ref{HO-wf}) and \protect(\ref{PL-wf}) are in $GeV$,
$f_{\rho}$ is $\rho$-meson leptonic decay constant in $MeV$.}
\label{tab:1}
\begin{tabular}{ccccccc}
\hline\hline\noalign{\smallskip}
~~Wave~~       &       &       &      &        &\\
~~functions~~  &$b\;$
               &$\langle\, r_\rho^2\rangle$
               &$\mu_\rho$& $Q_\rho$ &$f_{\rho}$\\
\noalign{\smallskip}\hline\noalign{\smallskip}
(\ref{HO-wf})    & 0.228 & 0.597 & 2.01 & -0.0064& 152.2\\
(\ref{PL-wf}) n=2& 0.217 & 0.579 & 2.12 & -0.0064& 153.6\\
(\ref{PL-wf}) n=3& 0.379 & 0.560 & 2.16 & -0.0066& 154.9\\
\noalign{\smallskip}\hline\hline
\end{tabular}
\end{table}

The results of calculations for the $\rho$-meson electromagnetic
form factors are represented in Fig.1--3.
\begin{figure}[h!]
\centering
\includegraphics[width=10.0cm]{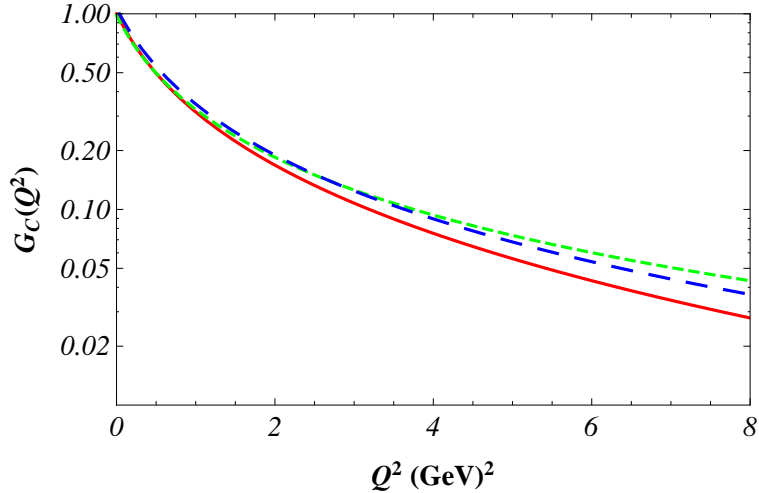}
\caption{The results of the calculations of the
$\rho$-meson charge form factor with different model wave
functions. The values of the parameters are given in the
Table 1. The solid line represents the relativistic calculation with the wave
function (\ref{HO-wf}), the short-dashed line -- with (\ref{PL-wf})
for $n =$ 2, the long-dashed line -- with (\ref{PL-wf})
for $n =$ 3.
\label{Plot:1}}
\end{figure}

\newpage

\begin{figure}[h!]
\centering
\includegraphics[width=10.0cm]{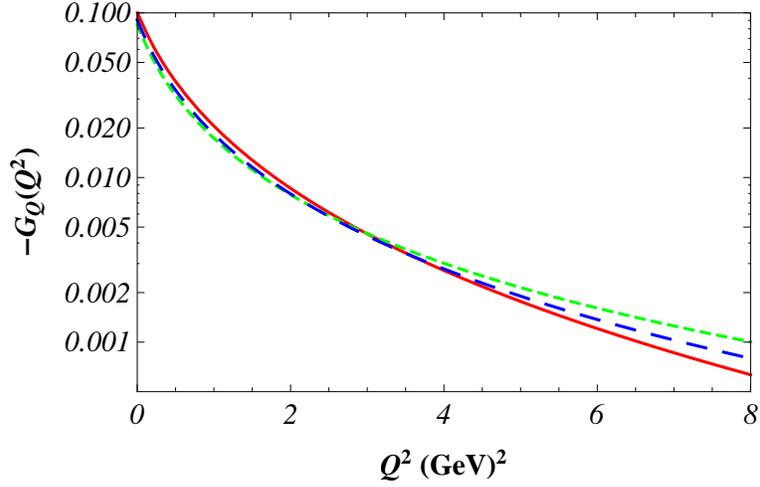}
\caption{The results of the calculations of the
$\rho$-meson quadrupole form factor with different model wave
functions, legend as in Figure 1}
\label{fig:2a}
\end{figure}

\begin{figure}[h!]
\centering
\includegraphics[width=10.0cm]{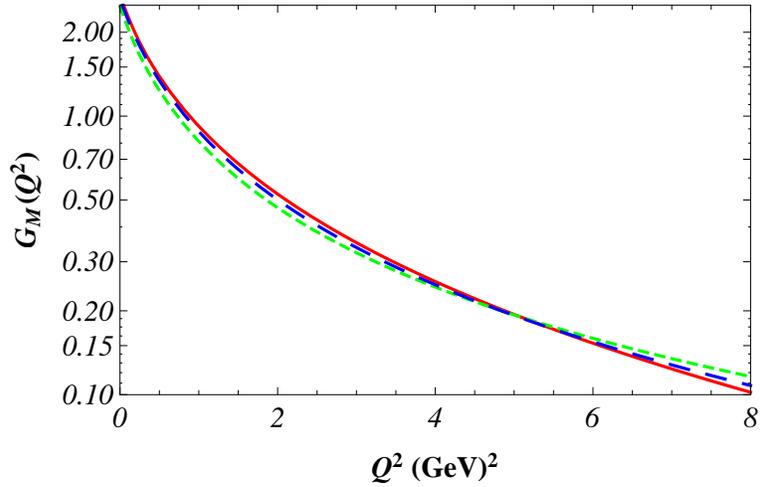}
\caption{The results of the calculations of the
$\rho$-meson magnetic form factor with different model wave
functions, legend as in Figure 1.
\label{Plot:3}}
\end{figure}

\section{Conclusion}

The electromagnetic form factors, quadrupole and magnetic moments,
MSR, and lepton decay constant of the $\rho$-meson were calculated
in the framework of the instant form of relativistic quantum
mechanic (RQM).

The special method of construction of the electromagnetic current
matrix elements for the relativistic two--particle composite
systems with nonzero total angular momentum is used to obtain the
integral representation for the electromagnetic form factors and
lepton decay constant.

The modified impulse approximation (MIA) is formulated in terms of
reduced matrix elements on Poincar\'e group. MIA conserves Lorentz
covariance of electroweak current and the electromagnetic current
conservation law.

A reasonable description of the electromagnetic static moments and
form factors and lepton decay constant of $\rho$-meson is obtained
in the developed formalism. Our approach gives the nonzero
quadrupole moment due to the relativistic Wigner spin rotation in
the $S$-state of the two quarks in $\rho$-meson.

So, it is shown that developed variant of the the instant form of
RQM can be used to obtain an adequate description of the
electroweak properties of composite systems with nonzero total
angular momentum.

}

\end{document}